\documentclass[10pt]{article}  
\usepackage{latexsym,amssymb}
\usepackage[dvips]{graphicx} 
\usepackage{enumerate}
\usepackage{amsmath,amssymb,amsfonts, amsthm}
\usepackage{graphicx}
\usepackage[dvips]{color}
\usepackage{epsf}

\renewcommand{\Vec}[1]{\mbox{\boldmath $#1$}}

\newcommand{\QED}{\hfill QED}
\newcommand{\Order}{\mbox{\rm O}}

\newtheorem{Lemma}{Lemma}
\newtheorem{Theorem}{Theorem}

\makeatletter
\def\Hline{%
\noalign{\ifnum0=`}\fi\hrule \@height 1.5pt \futurelet
\reserved@a\@xhline}
\makeatother

\begin{document}


\begin{center}
{\LARGE Characterizing a Set of Popular Matchings \\ 
   Defined by Preference Lists with Ties }\\
\vspace{10mm}

{\large
\begin{tabular}{ccc}
	Tomomi Matsui and  Takayoshi Hamaguchi \\ \\
	Graduate School of Decision Science and Technology,\\
	Tokyo Institute of Technology, \\
	Ookayama 2-12-1, Meguro-ku, Tokyo 152-8552, Japan \\ \\
   {January 13, 2016} \\
\end{tabular}
}

\begin{quote}
{\bf Abstract} 
In this paper, 
	we give a simple characterization 
	of a set of popular matchings
	defined by preference lists with ties.
By employing our characterization, 
	we propose a polynomial time algorithm 
	for finding a minimum cost popular matching.
\end{quote}

\end{center}

\renewcommand{\arraystretch}{1.2}


\section{Introduction}
In this paper, we give a characterization of a set of popular matchings
 in a bipartite graph with one-sided preference lists.
The concept of a popular matching was first introduced 
	by Gardenfors~\cite{GARDENFORS1975}.
Recently, Abraham et al.~\cite{ABRAHAM2007} 
	discussed a problem for finding a popular matching
	and proposed polynomial time algorithms 
	for problem instances defined by preference lists 
	with or without ties.
McDermid and Irving~\cite{MCDERMID2011} 
	discussed a set of popular matchings
	defined by strict preference lists.
One of a remained open problems raised 
	in~\cite{KAVITHA2009,MCDERMID2011} 
	and~\cite{MANLOVE2013} (Section~7.7)
	is a characterization of 
	a set of popular matchings when given preference lists have ties.
This paper solves the above open problem affirmatively
	and gives an explicit characterization of
	a set of popular matchings defined by preference lists with ties.
By employing our characterization, 
	we can transform a minimum cost popular matching problem, 
	discussed in~\cite{KAVITHA2009,MCDERMID2011}, 
	to a simple minimum cost assignment problem.

\section{Main Result}
An instance of popular matching problem comprises 
	a set $A$ of {\em applicants} and a set $P$ of {\em posts}.
Each applicant $a \in A$ has a preference list
	in which she ranks some posts in weak order
	(i.e., ties are allowed).
Given any applicant $a \in A$, and given any posts 
	$p, p' \in P$,
	applicant $a$ {\em prefers} $p$ to $p'$
	if both $p$ and $p'$ appear in $a$'s preference list,
	and $p$ precedes $p'$ on $a$'s preference list.
We assume that each applicant $a \in A$ has a specified post $l(a)$,
	called {\em last resort} of $a$, such that
	$l(a)$ appears only in $a$'s preference list and
	$l(a)$ is a unique, most undesirable post of $a$.
The existence of last resorts implies that $|A| \leq |P|$.
We say that a pair $(a, p) \in A \times P$ is {\em acceptable}
	if and only if post $p$ appears in $a$'s preference list.
We denote the set of acceptable pairs by $E \subseteq A \times P$. 
This paper deals with a bipartite graph
	$G=(A \cup P, E)$ consisting of vertex set $A \cup P$
	and edge set $E$.
Throughout this paper, we denote $|A|+|P|$ by $n$ and $|E|$ by $m$.

A subset of acceptable pairs $M \subseteq E$
	is called a {\em matching} if and only if 
	each applicant and post appears in at most one pair in $M$. 
We say that 
	$v \in A \cup P$ is {\em matched} in $M$, 
	when $M$ includes a pair $(a, p)$
	satisfying $v \in \{a, p\}$.
A set of matched nodes of $M$ is denoted by $\partial M$.
For each pair $(a, p) \in M$,  
	we denote $M(a)=p$ and $M(p)=a$. 

We say that an applicant $a \in A$ prefers matching $M'$ to $M$, 
	if (i) $a$ is matched in $M'$ and unmatched in $M$, or
	(ii) $a$ is matched in both $M'$ and $M$,
	and $a$ prefers $M'(a)$ to $M(a)$.
A matching $M'$ is {\em more popular} than $M$ 
	if the number of applicants that prefer $M'$ to $M$ exceeds
	the number of applicants that prefer $M$ to $M'$.
A matching $M$ is {\em popular} if and only if 
	there is no matching $M'$ that is more popular than $M$.
The existence of the set of last resorts implies that 
	we only need to consider applicant-complete matching, 
	since any unmatched applicant can be allocated to her last resort.
Throughout this paper, 
	we deal with popular matchings which are applicant-complete.
	
For each applicant $a \in A$, 
	we define $f(a) \subseteq P$ to be the set of 
	$a$'s most-preferred posts.
We call any such post $p \in f(a)$ an {\em f-post} of applicant $a$.
We define the {\em first-choice graph} of $G$ as
	$G_1=(A \cup P, E_1)$,
	where $E_1=\{(a, p) \in E \mid \exists a \in A, \; p \in f(a) \}$.


Let ${\cal M}_1$ be the set of maximum cardinality matchings of $G_1$
	and $k^*_1$ be the size of 
	a maximum cardinality matching of $G_1$.
Let $P_1 \subseteq P$ be a subset of posts   
	which are matched in every matching in ${\cal M}_1$.
For each applicant $a \in A$, 
	define $s(a)$ to be the set of
	most-preferred posts in $a$'s preference list 
	that are not in $P_1$.
Every post in $s(a)$ is called 
	an {\em s-post} of applicant $a$.
We define 
	$E_2=\{(a,p) \in A \times P \mid 
		p \in f(a) \cup s(a) \}$ and
	$G_2=(A \cup P, E_2)$.
	
Abraham et al.~\cite{ABRAHAM2007} showed 
	the following characterization of popular matchings.

\begin{Theorem}[{\rm Abraham et al.~}\cite{ABRAHAM2007}]\label{Aiff}
An applicant-complete matching $M \subseteq E$ of $G$ is popular if and only if 
	$M$ satisfies
 \mbox{\rm (i)}~$M \cap E_1 \in {\cal M}_1$  and
 \mbox{\rm (ii)}~$M \subseteq E_2$.
\end{Theorem}

Now. we describe our main result.
A {\em cover} of a given graph $G_1=(A \cup P, E_1)$ 
	is a subset of vertices $X \subseteq A \cup P$
	satisfying that $\forall (a, p) \in E_1$, 
	$\{a,p\} \cap X \neq \emptyset$.

\begin{Theorem}\label{MainTh1}
Assume that a given instance has at least one popular matching.
Let $X \subseteq A \cup P$ be a minimum cover of $G_1$.
We define	$\widetilde{P}=P \cap X$ and 
\begin{eqnarray*}
\widetilde{E}	
&=& 		\biggl( E_1 \cap (X_A \times \overline{X_P}) \biggr)
	\cup 	\biggl( E_1 \cap (\overline{X_A} \times X_P) \biggr)
	\cup 	\biggl( E_2 \cap \left( \overline{X_A} \times \overline{X_P}\right )\biggr)
\end{eqnarray*}
	where $X_A = A \cap X$, $X_P=P \cap X$, 
		$\overline{X_A} =A \setminus X$, and 
		$\overline{X_P} =P \setminus X$.
Then, an applicant complete matching $M$ in $G$ is popular if and only if
	$M \subseteq \widetilde{E}$ and 
	every post in $\widetilde{P}$ is matched in $M$.
\end{Theorem}

\section{Characterization of a Set of Popular Matching}


First, we introduce a maximum weight matching problem on $G_2$ defined by 
\begin{eqnarray}
\mbox{MP}: \mbox{maximize} 
&\mbox{\quad}&	|A| \sum_{e \in E_1} x(e)+ \sum_{e \in E_2} x(e) \nonumber \\ 
\mbox{subject to}
&&	
	\sum_{e \in \delta_2(a)}x(e) \leq 1 \;\;\; (\forall a \in A),\nonumber \\ 
&&	\sum_{e \in \delta_2(p)}x(e) \leq 1 \;\;\; (\forall p \in P), \nonumber \\ 
&&	x(e) \in \{0,1\} 	\;\;\; (\forall e \in E_2), \label{0-1const}
\end{eqnarray}
	where $\delta_2(v) \subseteq E_2$ denotes a set of edges 
	incident to a vertex $v \in A \cup P$ on $G_2$. 

\begin{Lemma}\label{optimality}
An applicant-complete matching $M \subseteq E$ is popular if and only if
	$M \subseteq E_2$ and 
	the corresponding characteristic vector 
	$\Vec{x} \in \{0,1\}^{E_2}$ defined by
\[ 
	x(e)=
	\left\{
		\begin{array}{ll}
			1 & (\mbox{\rm if } e \in M), \\
			0 & (\mbox{\rm otherwise}), \\
		\end{array}
	\right.
\]
	is optimal to \mbox{\rm MP}\@.
\end{Lemma}

\noindent
{\bf Proof.}
First, consider a case that a given 
	applicant-complete matching $M$ is popular.
Theorem~\ref{Aiff} states that $M \subseteq E_2$ 
	and $M \cap E_1 \in {\cal M}_1$.
Thus, we only need to show that 
	the characteristic vector 
	$\Vec{x}$ is optimal to MP\@.
The corresponding objective function value is equal to 
\[
	|A|\sum_{e \in E_1}x(e) + \sum_{e \in E_2}x(e)
	=|A|k^*_1 + |A|  
\]
	where $k^*_1$ denotes the size of 
	a maximum cardinality matching of $G_1$.
We show that the optimal value of MP is less than or equal to 
	$|A|k^*_1+|A|$. 
Let $\Vec{x}'$ be a feasible solution of MP and
	$M'=\{e  \in E_2 \mid x'(e)=1\}$.
Since $M' \cap E_1$ is a matching of $G_1$, 
	the objective function value corresponding to $\Vec{x}'$
	satisfies 
\[
	|A|\sum_{e \in E_1}x'(e)+\sum_{e \in E_2}x'(e)
	=|A| |M' \cap E_1| + |M'|
	\leq |A|k^*_1 + |A|  
\]
	which gives an upper bound of the optimal value of MP\@.
Thus, $\Vec{x}$ is optimal to MP\@.

Next, we consider a case that
	$M \subseteq E_2$ and 
	the corresponding characteristic vector $\Vec{x}$ of $M$
	is optimal to MP\@. 
Obviously, we only need to show that $M \cap E_1 \in {\cal M}_1$.
Assume on the contrary that	$M \cap E_1 \not \in {\cal M}_1$.
Let $M^* \in {\cal M}_1$ be a maximum cardinality matching of $G_1$
	and put $\Vec{x}^* \in \{0,1\}^{E_2}$ be the corresponding 
	characteristic vector.
The above assumption implies that $|M \cap E_1|+1 \leq |M^*|$. 
Obviously, $\Vec{x}^*$ is feasible to MP and satisfies
\begin{eqnarray*}
\lefteqn{
	|A|\sum_{e \in E_1}x^*(e) + \sum_{e \in E_2}x^*(e) 
	= |A||M^*|+|M^*|=(|A|+1)|M^*| 
}\\
	&\geq & (|A|+1) (|M \cap E_1|+1) 
	> |A||M \cap E_1|+|A| \\
	&= & |A||M \cap E_1|+|M| 
	= |A|\sum_{e \in E_1}x(e) + \sum_{e \in E_2}x(e),
\end{eqnarray*}
	which contradicts with the optimality of $\Vec{x}$.
From the above, we obtain that $M \cap E_1 \in {\cal M}_1$.
Theorem~\ref{Aiff} implies that $M$ is popular. 
\QED

\bigskip

Now we introduce a linear relaxation problem (LRP) of MP\@,
	which is obtained from MP by substituting 
	non-negative constraints
	$ x(e)\geq 0 \;\; (\forall e \in E_2)$
	for  0-1 constraints~(\ref{0-1const}).
It is well-known that every (0-1 valued) optimal solution
	of MP is also optimal to LRP~\cite{DANTZIG1951}.
A corresponding dual problem is given by
\begin{eqnarray*}
 \mbox{\rm D: minimize}& \quad 
 & 	\sum_{a \in A} y(a) + \sum_{p \in P} y(p) \\
 \mbox{\rm subject to} 
&& \begin{array}[t]{ll}
	y(a) + y(p) \geq |A|+1 	&(\forall (a, p) \in E_1), \\
	y(a) + y(p) \geq 1 		&(\forall (a, p) \in E_2 \setminus E_1),\\
    y(v) \geq 0 			&(\forall v \in A \cup P).
	\end{array}
\end{eqnarray*}

\begin{Theorem}\label{mainTh}
Let $\Vec{y}^*$ be an optimal solution of D.
We define
	$\widetilde{P}=\{p \in P \mid y^*(p)>0 \}$
	and 
\[
	\widetilde{E}=	\{ (a, p) \in E_1 \mid y^*(a)+y^*(p)=|A|+1\} 
			\cup	\{ (a, p) \in E_2 \setminus E_1 \mid 
											y^*(a)+y^*(p)=1 \}.
\]
An applicant-complete matching $M \subseteq E$ is popular 
	if and only if 
	\mbox{\rm (i)}~$ M \subseteq \widetilde{E}$, and 
	\mbox{\rm (ii)}~every posts in $\widetilde{P}$ is matched in $M$.
\end{Theorem}

\noindent
{\bf Proof.} 
Let $M \subseteq E$ be an applicant-complete matching 
	satisfying (i) and (ii).
Clearly, (i) implies that 
	$M \subseteq \widetilde{E} \subseteq E_2.$
From property (ii), 
	every post $p$ unmatched in $M$ satisfies that 
	$y^*(p)=0$.
The characteristic vector $\Vec{x}$ of $M$
	indexed by $E_2$ satisfies that
\begin{eqnarray*}
\lefteqn{
|A|\sum_{e \in E_1}x(e)+\sum_{e \in E_2}x(e) 
= (|A|+1) \sum_{e \in E_1} x(e) + \sum_{e \in E_2 \setminus E_1} x(e) } \\
&=& \sum_{e \in M \cap E_1}(|A|+1)  + \sum_{e \in M \cap (E_2 \setminus E_1)} 1\\
&=& \sum_{(a, p) \in M \cap E_1}(y^*(a)+y^*(p))
  + \sum_{(a, p) \in M \cap (E_2 \setminus E_1)} (y^*(a) + y^*(p) )\\
&=& \sum_{v \in \partial M} y^*(v) 
= \sum_{a \in A} y^*(a)+ \sum_{p \in P \cap \partial M}y^*(p)
= \sum_{a \in A} y^*(a) + \sum_{p \in P} y^*(p), 
\end{eqnarray*}  
where $\partial M$ denotes a set of vertices of $G_2$ matched in $M$.
Since $\Vec{x}$ is feasible to LRP\@,
	the weak duality theorem implies that 
	$\Vec{x}$ is optimal to LRP\@. 
	Clearly, $\Vec{x}$ is feasible to MP\@, 
	$\Vec{x}$ is also optimal to MP\@. 
Then, Lemma~\ref{optimality} implies the popularity of $M$.

Conversely, consider a case that
	an applicant-complete matching $M \subseteq E$ is popular.
Lemma~\ref{optimality} implies that
	$M \subseteq E_2$ and 
	the corresponding characteristic vector $\Vec{x}$ is optimal to MP\@. 
Dantzig~\cite{DANTZIG1951} showed that $\Vec{x}$ is also optimal to LRP\@.
The strong duality theorem implies that
\begin{align}
0 &= \left( \sum_{a \in A} y^*(a) + \sum_{p \in P} y^*(p) \right)
	- \left( |A|\sum_{e \in E_1}x(e)+\sum_{e \in E_2}x(e)  \right) 
	\nonumber \\
&=	\sum_{(a, p) \in M} (y^*(a) + y^*(p)) 
	+ \sum_{p \in P \setminus \partial M}y^*(p)
	- \left(
		(|A|+1) \sum_{e \in E_1} x(e) + \sum_{e \in E_2 \setminus E_1} x(e)
	\right) 
	\nonumber \\
&= \sum_{(a, p) \in M \cap E_1}(y^*(a)+y^*(p)-(|A|+1)) 
	\label{complement1} \\
&   + \sum_{(a, p) \in M \cap (E_2 \setminus E_1)} (y^*(a) + y^*(p) -1)
	+ \sum_{p \in P \setminus \partial M}y^*(p). \label{complement2} 
\end{align}
Since  $\Vec{y}^*$ is feasible to D, 
	each term in~(\ref{complement1}) or~(\ref{complement2}) is equal to $0$, 
	i.e., we obtain that
\begin{eqnarray*}
 &&	y^*(a)+y^*(p)=|A|+1, \;\; \forall (a,p) \in  M \cap E_1, \\ 
 && y^*(a)+y^*(p)=1, \;\; \forall (a,p) \in M \cap  (E_2 \setminus E_1), \\
 && y^*(p)= 0, \;\; \forall p \in  P \setminus \partial M.
\end{eqnarray*}
As a consequence, conditions (i) and (ii) hold.
\QED

\smallskip

Now we prove Theorem~\ref{MainTh1}.
Let us recall the following well-known theorem.
\begin{Theorem}[{\rm K\"onig} \cite{KONIG1931}]
The size of a minimum cover of $G_1$ is equal 
	the size of a maximum cardinality matching in $G_1$. 
\end{Theorem}

When we have a minimum cover, 
	we can construct an optima solution of dual problem D easily.
	
\begin{Lemma}
Assume that a given instance has at least one popular matching.
Let $X \subseteq A \cup P$ be a minimum cover of $G_1$.
When we define a vector $\Vec{y}^* \in \mathbb{Z}^{A \cup P}$ by
\begin{equation}\label{dualopt}
	y^*(v)=
	\left\{ 
		\begin{array}{ll}
		|A|+1 	& (\mbox{\rm if } v \in A \cap X), \\
		1		& (\mbox{\rm if } v \in A \setminus X), \\
		|A|		& (\mbox{\rm if } v \in P \cap X), \\
		0		& (\mbox{\rm if } v \in P \setminus X),
		\end{array}
	\right.
\end{equation}
	then $\Vec{y}^*$ is optimal to \mbox{\rm D}\@.
\end{Lemma}

\noindent
{\bf Proof.}
First, we show that $\Vec{y}^*$ is feasible to \mbox{\rm D}\@.
Obviously,  $\Vec{y}^*$ satisfies non-negative constraints.
For any edge $(a, p) \in E_2 \setminus E_1$, 
	$y^*(a)+y^*(p) \geq y^*(a) \geq 1$.
For any edge $(a, p) \in E_1$, 
	the definition of a cover implies that
	$\{a, p\} \cap X \neq \emptyset$, 	 
	and thus $y^*(a)+y^*(p) \geq |A|+1$.
From the above discussion, $\Vec{y}^*$ 
	is a feasible solution of D\@. 

Since there exists a popular matching, 
	the optimal value of MP is equal to
	$|A|k^*_1+|A|$.
K\"onig's theorem says that 
	$|X|=k^*_1$ and thus 
	the optimal value of MP is equal to 
	 $|A|k^*_1+|A|=|A||X|+|A|$.
The weak duality implies that
	$|A||X|+|A|$ gives a lower bound of 
	the optimal value of D\@.
Since $\Vec{y}^*$ is feasible to D
	and the corresponding objective value 
	attains the lower bound $|A||X|+|A|$,
	$\Vec{y}^*$ is optimal to D\@.
\QED

\smallskip

Lastly, we describe a proof of our main result.

\noindent
{\bf Proof of Theorem~\ref{MainTh1}}.
When a given instance has at least one popular matching, 
	dual solution $\Vec{y}^*$ defined by~(\ref{dualopt})
	is optimal to D\@.
Thus, Theorem~\ref{mainTh} directly implies Theorem~\ref{MainTh1}.
\QED

\smallskip

In the rest of this section, 
	we describe a method for constructing sets
	$\widetilde{P}$ and $\widetilde{E}$ efficiently.
First, we apply Hopcroft and Karp's algorithm~\cite{HOPCROFT1973}
	to $G_1$
	and find a maximum cardinality matching and 
	a minimum (size) cover $X$ of $G_1$ simultaneously.
Next, we employ an algorithm 
	proposed by Abraham et al.~\cite{ABRAHAM2007} and
	check the existence of a popular matching.
If a given instance has at least one popular matching, 
	then we construct sets $\widetilde{P}$ and $\widetilde{E}$
	defined in Theorem~\ref{MainTh1}.
The total computational effort required in the above procedure
	is bounded by $\Order (\sqrt{n}m)$ time.

\section{Optimal Popular Matching}

Kavitha and Nasre~\cite{KAVITHA2009} studied some problems 
	for finding a matching that is not only popular, 
	but is also optimal with respect to
	some additional criterion. 
McDermid and Irving~\cite{MCDERMID2011}
	discussed these problems 
	in case that given preference lists are strictly ordered,
	and proposed efficient algorithms based 
	on a specified structure called ``{\em switching graph}.''
Their algorithms find
(P1) a maximum cardinality popular matching 
		in $\Order (n+m)$ time, 
(P2) a minimum cost maximum cardinality popular matching 
		in $\Order (n+m)$ time, 
(P3) a rank-maximal popular matching
		in $\Order (n \log n+m)$ time, or 
(P4) a fair popular matching 
		in $\Order (n \log n+m)$ time.
They also showed that all the above problems are
	reduced to minimum weight popular matching problems
	by introducing an appropreate edge-cost $w: E_2 \rightarrow \mathbb{Z}$.

In the following, we discuss a minimum cost popular matching problem
	defined by preference lists with ties.
As shown in Theorem~\ref{MainTh1}, 
	we can characterize a set of popular matchings
	by a pair of sets
	$\widetilde{P}$ and $\widetilde{E}$.
Thus, we can find a minimum cost popular matching 
	by solving the following minimum cost assignment problem
\begin{eqnarray*}
\mbox{MCP:  minimize}&& \sum_{e \in E_2} w(e)x(e) \\
\mbox{subject to}
&& \sum_{e \in \delta_2 (a)}x(e)=1 		\;\; (\forall a \in A),\\
&& \sum_{e \in \delta_2 (p)}x(e)=1 		\;\; (\forall p \in \widetilde{P}),\\
&& \sum_{e \in \delta_2 (p)}x(e)\leq 1  \;\; (\forall p \in P \setminus \widetilde{P}),\\
&& x(e) =0 \;\;\; (\forall e \in E_2 \setminus \widetilde{E}). \\
&& x(e) \in \{0,1\} \;\;\; (\forall e \in E_2).
\end{eqnarray*}

\noindent
A well-known succesive shortest path method solves the above problem
	in $\Order (n(n \log n +m))$ time (see~\cite{AHUJA1993} for example).

\section{Discussions}

In this paper, 
	we give a simple characterization 
	of a set of popular matchings
	defined by preference lists with ties.
By employing our characterization, 
	we can find a minimum cost popular matching
	in $\Order (n(n \log n+ m))$ time.

When we deal with a problem for finding a popular matching 
	with a property (P1), (P2), (P3) or (P4), 
	there exists a possibility to reduce the time complexity,
	since the corresponding edge cost has a special structure.
However, we need a detailed discussion,
	since problem MCP has both equality and inequality constraints. 

We can construct an algorithm for enumerating all the popular matchings
	by employing an idea appearing in~\cite{FUKUDA1994}.
The required computational effort is bounded by 
	$\Order (\sqrt{n}m+ K(n+m))$
	where $K$ denotes the total number of popular matchings.
We omit the details of the enumeration algorithm.


\end{document}